\documentclass{sf2a-conf2011}
\usepackage{graphicx}
\usepackage{hyperref}
\usepackage[]{natbib}  
\usepackage[cyr]{aeguill}
\usepackage{epstopdf}
\usepackage{amssymb}

\def\BibTeX{{\rm B\kern-.05em{\sc i\kern-.025em b}\kern-.08em
    T\kern-.1667em\lower.7ex\hbox{E}\kern-.125emX}}
\bibpunct{(}{)}{;}{a}{}{,}  

\DeclareTextSymbol{\degre}{T1}{6}

\begin{document}

\TitreGlobal{SF2A 2011}


\title{Towards constraining the central black hole's properties by studying its infrared flares with the GRAVITY instrument}

\runningtitle{Constraining Sgr~A*'s properties with GRAVITY}

\address{LESIA, Observatoire de Paris, CNRS,  Universit\'e Pierre et Marie Curie, Universit\'e  Paris Diderot, 5 place Jules Janssen, 92190 Meudon, France}
\address{LUTH, Observatoire de Paris, CNRS, Universit\'e Paris Diderot, 5 place Jules Janssen, 92190 Meudon, France}
\address{Max-Planck-Institut f\"ur Extraterrestrische Physik, 85748 Garching, Germany}
\author{F. H. Vincent$^{1,2}$}

\author{T. Paumard$^{1}$}

\author{G. Perrin$^{1}$}

\author{E. Gourgoulhon$^{2}$}

\author{F. Eisenhauer$^{3}$}

\author{S. Gillessen$^{3}$}


\setcounter{page}{237}

\index{Vincent, F. H.}
\index{Paumard, T.}
\index{Perrin, G.}
\index{Gourgoulhon, E.}


\maketitle


\begin{abstract}
The ability of the near future second generation VLTI instrument GRAVITY to constrain the properties of the Galactic center black hole is investigated. The Galactic center infrared flares are used as probes of strong-field gravity, within the framework of the hot spot model according to which the flares are the signature of a blob of gas orbiting close to the black hole's innermost stable circular orbit. Full general relativistic computations are performed, together with realistic observed data simulations, that lead to conclude that GRAVITY could be able to constrain the black hole's inclination parameter.
\end{abstract}



\section{Introduction}
The existence of a supermassive black hole coincident with the radio source Sgr~A* at the center of our Galaxy is highly probable, as advocated by decades of observations~\citep[see][for the most recent]{ghez08,gillessen09}. Moreover, Sgr~A* exhibits outbursts of radiation, hereafter flares, in the millimeter, infrared and X-ray wavelengths~\citep[see][and references therein, for the most recent observations]{trap11}. No consensus has yet been reached regarding the physical nature of these flares. Different models have been proposed in the literature: adiabatic expansion of a synchrotron-emitting blob of plasma \citep{yusefz06}, heating of electrons in a jet \citep{markoff01}, Rossby wave instability in the disk \citep{tagger06}, or a clump of matter heated by magnetic reconnection orbiting close to the innermost stable circular orbit (ISCO) of the black hole \citep{hamaus09}. To be tested, this last model, hereafter hot spot model, requires an astrometric precision at least of the order of the angular radius of the ISCO, which is a few times the Schwarzschild radius of the black hole, i.e. a few times $10 \, \mu \mathrm{as}$.

Such a precision will be within reach of the near future GRAVITY instrument~\citep{eisenhauer08}. \citet{vincent11} have shown that GRAVITY will be able of putting in light the motion of a spot orbiting on the ISCO of a Schwarzschild black hole, however without taking into account any relativistic effect. The aim of this paper is to determine, in the framework of a full general relativistic treatment, the ability of GRAVITY to get information on the properties of the central black hole by giving access to the astrometry of near infrared flares.

\section{Observing a hot spot orbiting around Sgr~A*}

The hot spot model used here is developed by~\citet{hamaus09}. The central black hole is assumed to be surrounded by a magnetized accretion disk. Due to differential rotation, the magnetic field lines are stretched, which leads to reconnection events that violently heat some part of the disk, giving rise to a hot sphere of synchrotron emitting plasma orbiting around the black hole on some circular orbit at a radius $r \gtrsim r_{\mathrm{ISCO}}$. The sphere is assumed to emit isotropically and to have a radius $r_{\mathrm{sph}}=0.5\,M$, where $M$ is the black hole's mass~\citep[this value is consistent with the constraint on the spot's radius given by][]{gillessen06}. Due to differential rotation, the sphere is distorted and finally forms an arc. The so-called hot spot is thus made of the superimposition of the initial sphere and of the arc resulting from its stretching. In order to take into account the heating and cooling phases due to the reconnection event, the sphere's and the arc's emission are multiplied by a gaussian temporal modulation. For the simulations, the arc is modeled as the sum of nearby spheres over one complete period around the black hole. Moreover, the arc's emitted intensity is modulated by an azimuthal gaussian that allows to peak the emission on the initial sphere's position. The parameters describing the hot spot are thus the standard deviations of these gaussian modulations, $\sigma_{\mathrm{\uparrow,sph}}$ and $\sigma_{\mathrm{\uparrow,arc}}$ for the heating modulations of the sphere and the arc, $\sigma_{\mathrm{\downarrow,sph}}$ and $\sigma_{\mathrm{\downarrow,arc}}$ for their cooling modulations, $\sigma_{\mathrm{azimuth}}$ for the arc's azimuthal modulation. One last parameter is added, the ratio $\rho_{I}$ of the sphere's mean luminosity to the arc's mean luminosity. This last parameter allows the bright sphere to dominate the emission. In all the following computations, these parameters will be maintained at the following values:
\begin{eqnarray}
  \label{paramflarestd}
  &&\sigma_{\mathrm{azimuth}} = 2\,\pi \\ \nonumber
  &&\rho_{I} = 60\\ \nonumber
  &&\sigma_{\mathrm{\uparrow,sph}} = 0.5\,t_{\mathrm{per}}, \quad \sigma_{\mathrm{\downarrow,sph}} = t_{\mathrm{per}} \\ \nonumber
  &&\sigma_{\mathrm{\uparrow,arc}} = 0.5\,t_{\mathrm{per}}, \quad\sigma_{\mathrm{\downarrow,arc}} = t_{\mathrm{per}}\\ \nonumber
\end{eqnarray}
where $t_{\mathrm{per}}$ is the period of the orbit. A general study taking into account the variations of these parameters goes beyond the scope of this work. However, the strongest hypothesis in the perspective of the following astrometric analysis is the big value of the parameter $\rho_{I}$. If this parameter is smaller, the sphere does not dominate the total emission, and the motion of the photocenter of the object is much attenuated, making it difficult, if not impossible, for an instrument to measure it. Nevertheless, let us note that the value of~$\rho_{I}$ fitted on observed data by~\citet{hamaus09} (see their Table 2) is even higher than the value used here.

\begin{figure}[ht!]
 \centering
 \includegraphics[width=0.48\textwidth,clip]{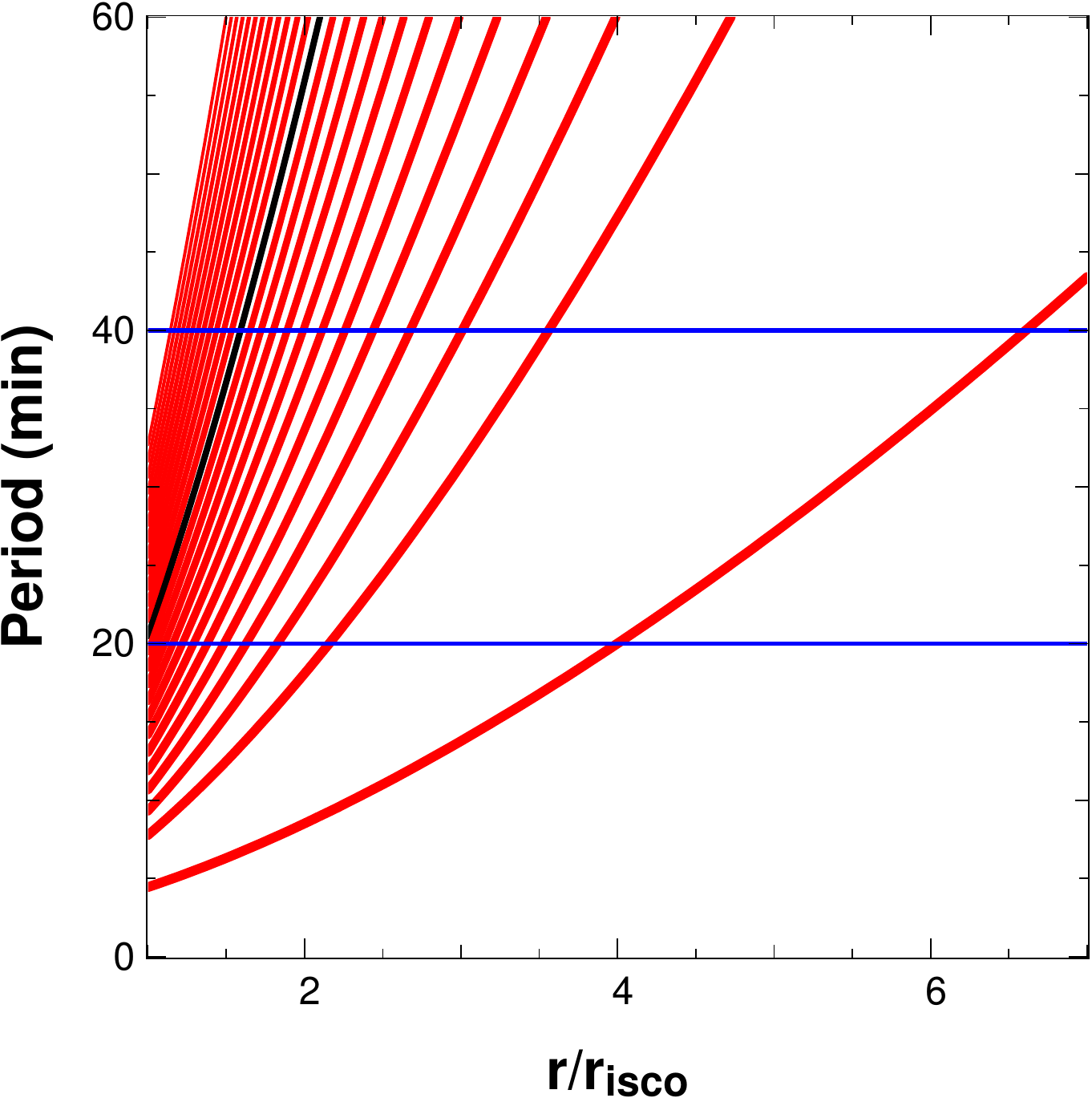}%
 \includegraphics[width=0.48\textwidth,clip]{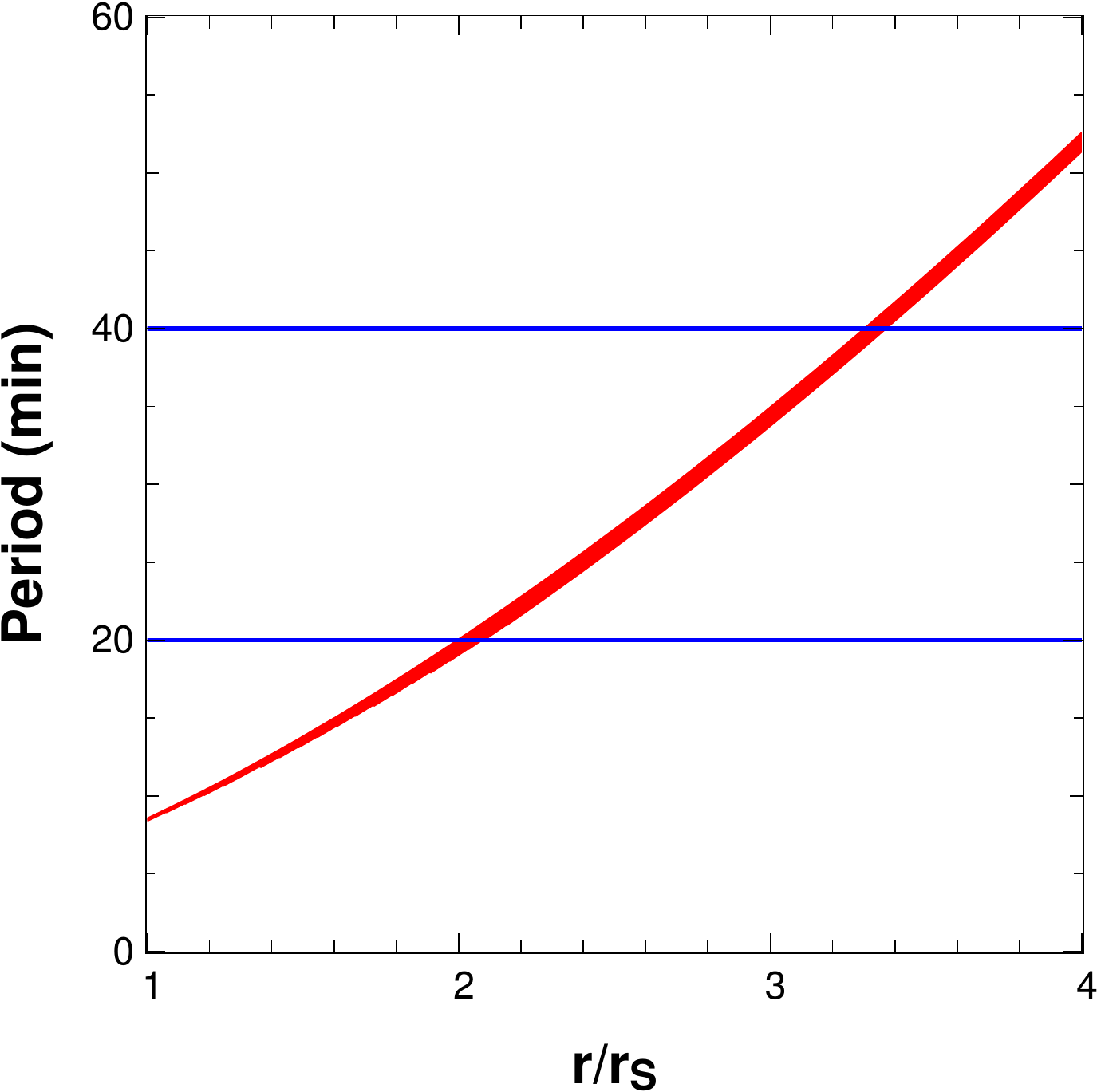}      
  \caption{{\bf Left:} Circular orbital period around a Kerr black hole of mass $4.31\,10^{6} M_{\odot}$ with spin parameter between 0 (top left curve) to 1 (down right curve), as a function of the orbital radius in units of the ISCO radius. The blue horizontal lines correspond to the range of observed pseudo periods: typically between $20$ and $40$~min.The black curve corresponds to spin $0.5$: spins $\lesssim 0.5$ are not able of accounting for the smallest observed pseudo period of flares of $\approx 20$~min. {\bf Right:} Same as left panel, but as a function of the orbital radius in units of the Schwarzschild radius of the black hole. The curves are truncated for values of $r$ smaller than the corresponding ISCO. }
  \label{vincent:fig1}
\end{figure}

The radius of the circular orbit of the hot spot and the black hole's spin parameter can be constrained by the near infrared observations of flares already obtained. These observations show a pseudo-periodic variability of the light curve that varies between around 20~min and around 40~min~\citep{genzel03,doddseden10}. In the framework of the hot spot model, these pseudo-periodicities correspond to the orbiting period of the spot around the black hole. Figs~\ref{vincent:fig1}, representing the evolution of the orbital period around the central black hole as a function of the radius normalized by the ISCO radius or the Schwarzschild radius $r_{\mathrm{S}}$, immediately lead to the conclusion that the black hole's spin parameter $a$ and hot spot's orbiting radius $r$ must satisfy:

\begin{eqnarray}
\label{range}
&a \gtrsim 0.5 , \\ \nonumber
&2\,r_{\mathrm{S}} \lesssim r \lesssim 3.5\,r_{\mathrm{S}}. \\ \nonumber
\end{eqnarray}

The computation of the hot spot trajectory around the black hole, as well as the simulation of its observed appearance for an observer on Earth is computed by means of the ray-tracing algorithm GYOTO\footnote{Freely available at the following URL: \url{http://gyoto.obspm.fr}}~\citep{vincent11b}.

\section{Towards constraining the inclination of the central black hole}

In this section, we wish to investigate the impact of the black hole's inclination on the GRAVITY astrometric simulated data. The impact of the spin parameter will not be investigated, and it is thus fixed to $a = 0.7\,M$, in agreement with the lower bound given in Eq.~\ref{range}. Moreover, the radius of the circular orbit of the hot spot is fixed to its highest possible value (see Eq.~\ref{range}): $r = 3.5\,r_{\mathrm{S}}$. One last parameter that has a strong impact on the astrometric data is the maximum magnitude of the hot spot. It is fixed to $m_{\mathrm{K}}=13$. These values are optimistic as the hot spot is very bright and evolves on a large trajectory. However, these values are not unrealistic as the brightest flare observed to date had a maximum magnitude of $m_{\mathrm{K}}=13.5$~\citep{doddseden11}, and the orbital radius of the hot spot investigated in Sect.~3.3.2 of~\citet{hamaus09} is fitted to a value close to $3.5\,r_{\mathrm{S}}$\footnote{Let us note that the central black hole's mass used in~\citet{hamaus09} differs from the mass used to obtain Figs~\ref{vincent:fig1} thus changing the timescale.}. Moreover, as will be stressed below, only one flare observed with such parameters would allow getting interesting information on the central black hole. These values are not assumed to be standard hot spot parameters.

\begin{figure}[ht!]
 \centering
 \includegraphics[width=0.48\textwidth,clip]{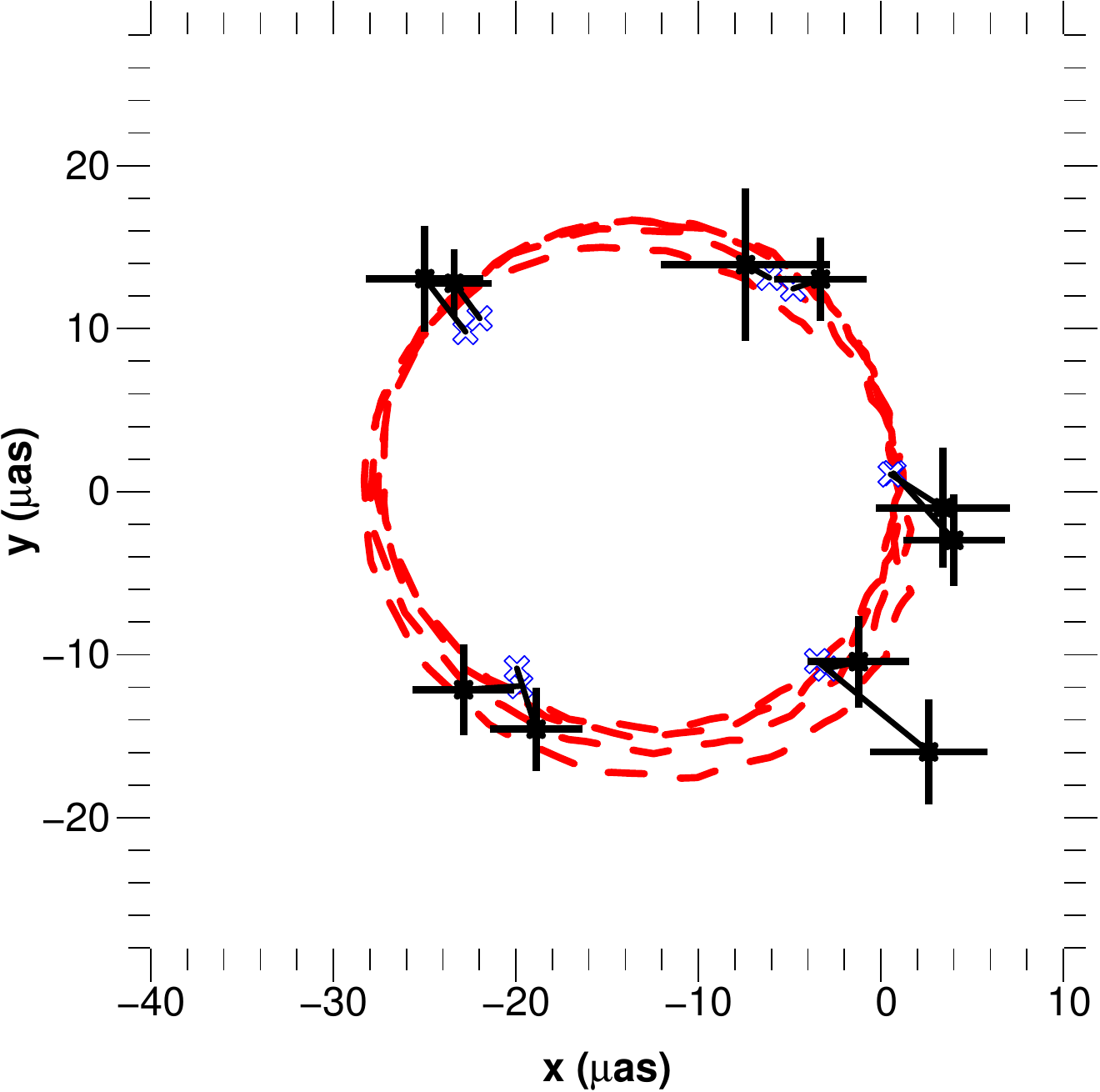}%
 \includegraphics[width=0.48\textwidth,clip]{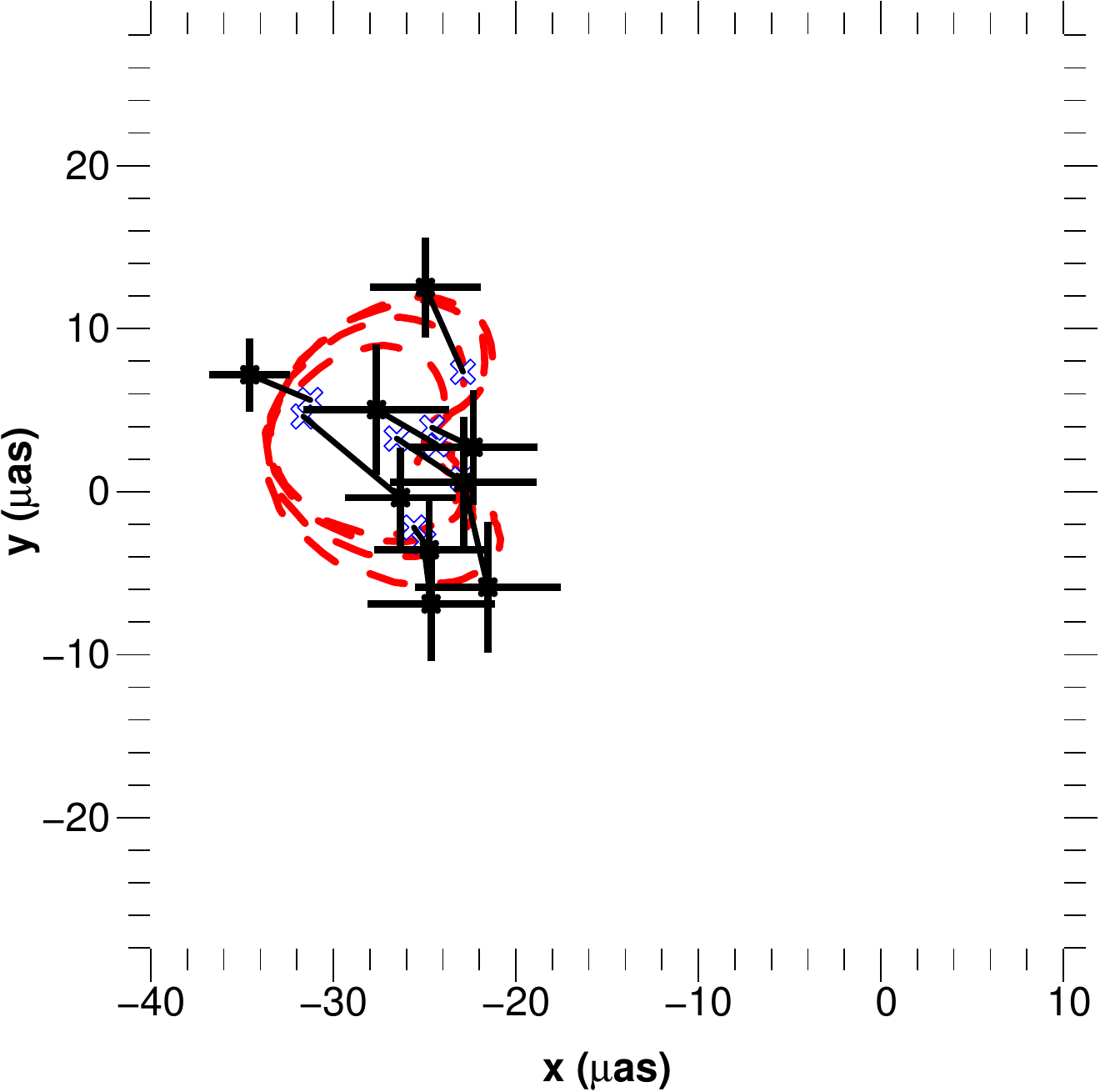}      
  \caption{{\bf Left:} GRAVITY observation simulation of a hot spot circularly orbiting at a radius $r=3.5\,r_{\mathrm{S}}$ around a Kerr black hole of spin parameter $0.7$ seen under an inclination of $20$~\degre~. The maximum magnitude of the flare is $m_{\mathrm{K}}=13$. The hot spot's parameters are given in Eq.~\ref{paramflarestd}. The red dashed line is the theoretical projected trajectory as computed by GYOTO. The black error bars are simulations of GRAVITY astrometric data (the size of the error bar is a function of the hot spot's magnitude at the time of observation), the instrument integrates during 100~s for each data point. Each black retrieved position is linked to the corresponding real position of the spot along its orbit (in blue). {\bf Right:} Same as left panel, but with an inclination parameter of $60$~\degre~. }
  \label{vincent:fig2}
\end{figure}

The theoretical trajectory of the hot spot projected on the observer's screen and the observed flux being known by using the GYOTO code, it is possible to simulate accurately the hot spot's astrometric positions that would be obtained by GRAVITY~\citep[see][for the description of this procedure]{vincent11}. Fig.~\ref{vincent:fig2} shows the simulation of such an observation by GRAVITY, with two different values of the inclination parameter. Let us stress that the error bars appearing on this figure take into account all sources of noise that will affect a real GRAVITY observation~\citep[see][]{vincent11}.

Fig.~\ref{vincent:fig2} shows clearly that the dispersion of the retrieved positions depends strongly on the inclination: the smaller the inclination, the bigger the dispersion. Building upon this result, we have investigated the ability to get an information on the inclination parameter of the black hole by computing the dispersion of the retrieved positions. This was done within the framework of a Monte Carlo analysis, by computing the dispersion values in the $x$ and $y$ directions corresponding to the horizontal and vertical directions of the observer's screen, for a number of simulated observations.

\begin{figure}[ht!]
 \centering
 \includegraphics[width=0.48\textwidth,clip]{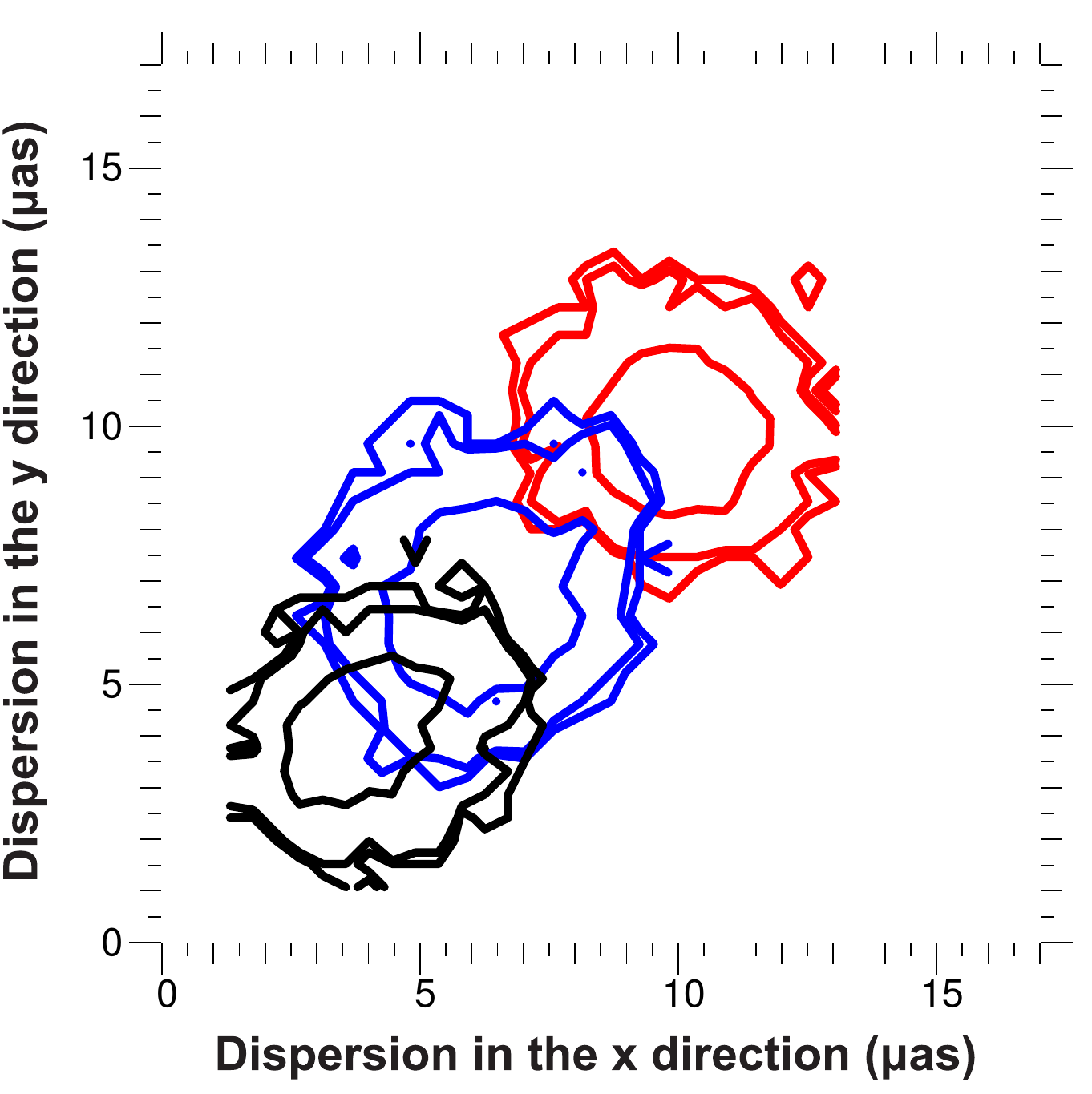}%
  \caption{2-dimensional histograms of the dispersion of the retrieved position in the two horizontal (x) and vertical (y) directions of the observer's screen. The hot spot parameters are the same as in Fig~\ref{vincent:fig2}, with an inclination parameter of $20$~\degre~(red), $40$~\degre~(blue) and $60$~\degre~(black). The contours include respectively (innermost to outermost) 68\%, 95\% and 99\% of the dispersion values obtained in the Monte Carlo procedure.}
  \label{vincent:fig3}
\end{figure}

Fig.~\ref{vincent:fig3} represents the bidimensional histograms of the retrieved positions in the $x$ and $y$ directions, for three different values of the inclination. The histograms corresponding to the three values of inclination, $20$~\degre~, $40$~\degre~and $60$~\degre~, span different values of the dispersion. Thus, if one considers one given night of observation satisfying the above assumption on the flare's maximum magnitude and orbital radius (which can be tightly constrained from the measured orbital period, see Fig.~\ref{vincent:fig1}, right panel), the inclination parameter can be constrained by computing the dispersion of the retrieved positions found by GRAVITY on this given night. For instance, if the dispersion in both direction is found to be $10\,\mu$as, Fig.~\ref{vincent:fig3} excludes at 3 $\sigma$ any inclination values $\gtrsim 40$~\degre~. This constraint could be refined by choosing a finer sampling of the inclination parameter.

\section{Conclusion}
Within the framework of the hot spot model for the Galactic center infrared flares, we have shown that, provided one bright enough ($m_{\mathrm{K}}=13$ at maximum) flare with a large enough ($r \approx 3.5\,r_{\mathrm{S}}$) orbital radius is observed, the GRAVITY instrument would be able of constraining the inclination parameter of the central black hole. As no consensus has yet emerged on the value of this important parameter, such a measure would allow progressing towards a better understanding of the innermost Galactic center.

\begin{acknowledgements}
This work was supported by grants from R\'egion Ile-de-France.
\end{acknowledgements}

\bibliographystyle{aa}  
\bibliography{vincent} 

\end{document}